\documentclass[reprint,amsmath,amssymb,aps,prl]{revtex4-2}
\usepackage{graphicx}
\usepackage{color}
\usepackage{hyperref}
\hypersetup{
    colorlinks=true,
    linkcolor=blue,
    citecolor=blue,
    filecolor=blue,
    urlcolor=blue,
}

\begin{document}
\title{One-dimensional non-Hermitian band structures as Riemann surfaces}
\author{Heming Wang$^{1,*}$, Lingling Fan$^{1,*}$ and Shanhui Fan$^{1,\dagger}$}
\affiliation{
$^1$Department of Electrical Engineering and Edward L. Ginzton Laboratory, Stanford University, Stanford, California 94305, USA\\
$^*$These authors contributed equally to this work.\\
$\dagger$Corresponding author: shanhui@stanford.edu}
\date{\today}

\begin{abstract}
We present the viewpoint of treating one-dimensional band structures as Riemann surfaces, linking the unique properties of non-Hermiticity to the geometry and topology of the Riemann surface.
Branch cuts and branch points play a significant role when this viewpoint is applied to both the open-boundary spectrum and the braiding structure.
An open-boundary spectrum is interpreted as branch cuts connecting certain branch points, and its consistency with the monodromy representation severely limits its possible morphology.
A braid word for the Brillouin zone can be read off from its intersections with branch cuts, and its crossing number is given by the winding number of the discriminant.
These results open new avenues to generate important insights into the physical behaviors of non-Hermitian systems.
\end{abstract}
\maketitle

{\it Introduction.}---Energy band theory, which relates the energies and wavevectors of the states in a physical system, provides the foundation for our understanding of solid-state physics \cite{ashcroft1976solid}, and underlies the recent developments in electromagnetic and acoustic metamaterials \cite{joannopoulos1997photonic, cummer2016controlling, ozawa2019topological}. While energy band theory was initially developed for Hermitian systems that satisfy energy conservation, its physics content is significantly enriched with the introduction of non-Hermiticity \cite{ashida2020non, bergholtz2021exceptional}.

In standard Hermitian energy band theory, where both the energies and the wavevectors are real, multiple bands that are separated by band gaps are usually treated as separate entities that are largely independent of each other. On the other hand, the development of non-Hermitian energy band theory necessitates the treatment of both the energies and the wavevectors as complex quantities. In this Letter, we propose that, with complex energies and wavevectors, the multiple bands need to be considered as subsets of a single object: a Riemann surface.

The treatment of band theory in terms of Riemann surfaces provides significant new insights into the topology of non-Hermitian bands. Experimentally, two important observable effects of one-dimensional (1D) non-Hermitian models are the open-boundary-condition (OBC) spectrum \cite{ghatak2020observation, helbig2020generalized, hofmann2020reciprocal, liu2021non, zhang2021acoustic} and the braiding relations of the bulk bands \cite{wang2021topological, patil2022measuring, tang2022experimental, zhang2023experimental}. The OBC spectrum \cite{yokomizo2019non, yang2020non, zhang2020correspondence, wu2022connections} exhibits a non-Hermitian skin effect \cite{yao2018edge, alvarez2018non, lee2019anatomy, kunst2019non, song2019nonB, longhi2019probing, okuma2020topological, yi2020non, kawabata2020non, kawabata2020higher, li2020critical, zhong2021nontrivial, zhang2022universal, zhang2022review}, which highlights a unique aspect of bulk-edge correspondence in non-Hermitian systems \cite{lee2016anomalous, leykam2017edge, shen2018topological, yin2018geometrical, kunst2018biorthogonal, yao2018non, xiong2018does, gong2018topological, zhou2019periodic, herviou2019defining, kawabata2019symmetry, edvardsson2019non, jin2019bulk, imura2019generalized, song2019nonA, borgnia2020non, zirnstein2021bulk, ding2022non}. On the other hand, the braiding of the bulk bands \cite{hu2021knots} provides a homotopic classification of non-Hermitian band structures \cite{wojcik2020homotopy, li2021homotopical} and has been observed in photonic \cite{wang2021topological, patil2022measuring} and acoustic \cite{tang2022experimental, zhang2023experimental} systems. Here we show that both the OBC spectrum and the braid crossings are related to the monodromy representation of the Riemann surface mapping. For the OBC spectrum, the monodromy representation strongly constrains its morphology. For the braiding in the bulk band structure, we express the crossing number as the winding number of the discriminant, which counts the branch points enclosed in the Brillouin zone. Our results highlight the geometric properties of the Riemann surface as defined by the band structure and their role in defining the physical aspects of non-Hermitian systems.

\begin{figure}
\centering
\includegraphics[width = 85mm]{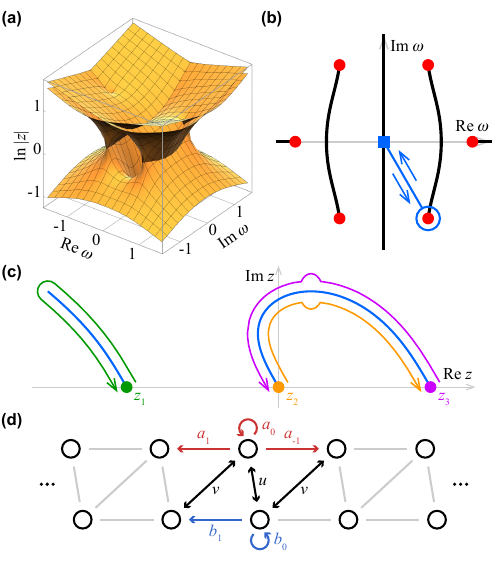}
\caption{Band structures as Riemann surfaces. Here $f$ is given by Eq. (\ref{eq:hexagon}).
(a) A 3D plot of $\ln |z|$ versus $\mathrm{Re}~\omega$ and $\mathrm{Im}~\omega$, showing the Riemann surface structure before projecting onto the $\omega$ plane.
(b) Branch cuts (black curves) and branch points (red points) of $f$ projected on the $\omega$ plane. The base point $\omega_0 = 0$ (blue square) and a based loop for the monodromy representation for one of the branch points are shown.
(c) As $\omega$ circles the branch point in (b), starting and finishing at the base point, some of the corresponding $z$'s (here $z_{(2)}$ and $z_{(3)}$) swap locations.
(d) A realization of $f$ as a two-band lattice model. Between the two chains, there are staggered couplings similar to the Su-Schrieffer-Heeger model \cite{su1979solitons}.
}
\label{fig:1}
\end{figure}

{\it Band structures as Riemann surfaces.}---We consider a tight-binding Hamiltonian on a 1D periodic structure with $r$ bands, with matrix elements
\begin{equation}
H_{mn}(k) = \sum_{-p \leq s \leq q}t_{mn,s}z^s.
\end{equation}
Here $t_{mn,s}$ is the complex hopping amplitude from band $n$ to band $m$ jumping $|s|$ cells to the left when $s>0$ and to the right when $s<0$, $p$ ($q$) are maximum coupling ranges to the right (left), $z=e^{ik}$, and $k\in\mathbb{C}$ is the wavevector. The characteristic equation $\mathrm{det}(H-\omega\mathbb{I})=0$ can be written as a polynomial $f(\omega, z) = 0$, which is $r$-th order in $\omega$ and $u$-th order in $z$ with $u \le (p+q)r$. The equation $f=0$ can be solved to obtain the band structure and the mapping between $\omega$ and $z$, where $r$ separate band frequencies can be found for a given $z$. We note that the band structure as defined by $f = 0$ includes both the bulk and the edge state behaviors. On the other hand, the equation $f=0$ can also be regarded as an {\it affine algebraic curve}, and can be converted into a Riemann surface after suitable compactification if $f$ is smooth \cite{cavalieri2016riemann}. The multivalued mapping of the Riemann surface as defined by $f = 0$ to the $\omega$ and $z$ complex planes may create branch points on these complex planes [Figs. \ref{fig:1}(a)-(b)]. Branch point locations in the $\omega$ plane can be solved from the discriminant of $f$ with respect to $z$, obtained by eliminating $z$ from $f=0$ and $\partial_z f=0$, and similarly for the $z$ plane. The number of branch points is governed by the Riemann-Hurwitz formula (see Supplementary Information \cite{SI}).

On the $\omega$ and $z$ complex planes, circling around one branch point will move the state from one band to another, and all bands could be connected in this way. This is captured by the {\it monodromy representation} \cite{cavalieri2016riemann}, which associates a permutation with each branch point identifying how the single-valued sheets are connected. On the $\omega$ plane, we label a specific value of $\omega$ different from the branch points as the base point $\omega_0$, which corresponds to a total of $u$ different $z$ values, arbitrarily labeled as $z_1$, $z_2$ $\cdots$ $z_u$ (recall that $u$ is the order of $z$ in the polynomial $f$). As $\omega$ continuously moves along a based loop, starting from the base point, circling a branch point, and eventually returning to the base point, the corresponding $z$'s may exchange locations, resulting in the permutation of the set $\{z_1 \cdots z_u\}$ [Fig. \ref{fig:1}(c)]. If $z_\alpha$ and $z_\beta$ swap places while the others return to their original positions, the permutation can be labeled as $(\alpha\beta)$ using the cycle notation. The permutation will depend on both the location of the base point and how the based loop winds around other branch points. For concreteness, here we always use a straight line segment connecting the base point and the branch point together with a small counterclockwise loop around the branch point, and without loss of generality, we assume that such a line segment does not cross another branch point. The based loops and permutations for all branch points collectively constitute the monodromy representation. 

Valid monodromy representations for band structures must be {\it consistent} and {\it connected}. Consistency requires that certain compositions of permutations of all branch points must be the identity. This can be seen by decomposing a small clockwise loop circling the base point as a sum of all the loops circling every branch point. On the other hand, every $z_\alpha$ can be reached by starting from any one of them and going through permutations, which is a consequence of the connectedness of the Riemann surface for $f$.

{\it The model.}---For illustration purposes, we will work with the following polynomial throughout the Letter:
\begin{equation}
f: z \omega^2 + \omega \sum_{s=0}^{2}A_s z^s + \sum_{s=0}^{3}B_s z^s = 0.
\label{eq:sixbp}
\end{equation}
This model is chosen because it is sufficient to illustrate the connection between the Riemann surface properties and the OBC spectrum and band braiding behaviors. A general $\omega$ corresponds to three $z$ values, and there are $6$ branch points on the $\omega$ plane. Similarly, a general $z$ corresponds to two $\omega$ values, and there are $4$ branch points on the $z$ plane. A band structure described by Eq. (\ref{eq:sixbp}) can be realized as a two-band lattice model [Fig. \ref{fig:1}(d)] with a Hamiltonian given by,
\begin{equation}
H = 
\begin{pmatrix}
a_1 z + a_0 + a_{-1}z^{-1} & u + v z^{-1} \\ 
v z + u & b_1 z + b_0
\end{pmatrix}.
\end{equation}
All coupling coefficients are complex numbers in general. With a particular choice of parameters (the parameters of this model, as well as all other models used in the Letter, are provided in Supplementary Information \cite{SI}), Eq. (\ref{eq:sixbp}) becomes,
\begin{equation}
z\omega^2 + \frac{1}{\sqrt[3]{2}}\left(z^2 - 1\right)\omega + \frac{1}{\sqrt[3]{4}}\left(\frac{z^3}{3} - z\right) = 0,
\label{eq:hexagon}
\end{equation}
with the $\omega$ branch points located on the six roots of unity. This choice of parameters will be used to demonstrate the connection between branch points and the physical observables.

\begin{figure}
\centering
\includegraphics[width = 85mm]{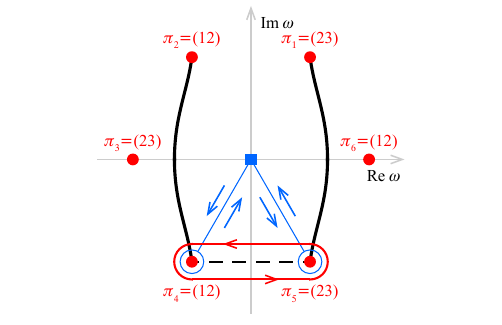}
\caption{The OBC spectrum for Eq. (\ref{eq:hexagon}), showing the branch points and the associated permutation labels. A branch cut along the black dashed line would be inconsistent with the monodromy representation, as seen by decomposing a small loop around the hypothetical branch cut (red) as a sum of two based loops (blue).
}
\label{fig:2}
\end{figure}

{\it OBC spectra.}---The continuous part of the OBC spectrum is defined by the generalized Brillouin zone $|z_{(\mu)}| = |z_{(\mu+1)}|$, where $z_{(l)}$ is the roots of $f=0$ for a specific $\omega$ sorted in increasing magnitude and $\mu$ is the number of boundary conditions on the left end of the chain \cite{yokomizo2019non, zhang2020correspondence}. By defining the $l$-th Riemann sheet to be composed of $\ln |z_{(l)}|$, the OBC spectrum becomes the branch cut associated with the $\mu$-th and $(\mu+1)$-th sheet on the $\omega$ plane. The OBC spectrum must be connected to the branch points where $z_{(\mu)}$ = $z_{(\mu+1)}$. Consequently, the monodromy representation for the branch points can be used to understand aspects of the morphology of the OBC spectrum. For example, suppose that two branch points are connected by a single branch cut and no other branch cuts are in the vicinity. A small loop around this branch cut will result in a trivial permutation. By decomposing this loop as a sum of based loops around the branch points, the permutation product formed from the labels for these based loops must also be the identity.

The OBC spectrum for Eq. (\ref{eq:hexagon}) is shown in Fig. \ref{fig:2}, consisting of two separate arcs that coincide with two of the branch cuts in Fig. \ref{fig:1}(b). We label the branch points by $s=1$, $2$ $\cdots$ $6$ according to their positions $\omega = \exp(2\pi i s/6)$, and their associated permutations by $\pi_1$ $\cdots$ $\pi_6$. The four branch points $1$, $2$, $4$ and $5$ satisfy $z_{(1)} = z_{(2)}$ and are included in the OBC spectrum (here $\mu = 1$), while the other two have $z_{(2)} = z_{(3)}$ and are not in the spectrum. Consistency with the monodromy representation then indicates that a pairwise vertical connection of the four branch points is possible instead of e.g. a pairwise horizontal connection. To see this, assume that a hypothetical branch cut connects branch points $4$ and $5$ with no other branch cuts in the vicinity [Fig. \ref{fig:2}]. The permutation along the loop around this hypothetical branch cut is given by $\pi_4\pi_5 = (132)$ and is nontrivial. As such, this branch cut could not be the only one originating from these two branch points. On the other hand, the consistency of the actual OBC spectrum can be readily verified as $\pi_1\pi_5 = \pi_4\pi_2 = 1$. We note that two branch points connected by a branch cut may have different permutation labels, an example of which is provided in Supplementary Information \cite{SI}. In general, with the knowledge of permutation labels for each branch point (which can be computed relatively straightforwardly), the possible forms of branch cuts, and in turn the OBC spectrum, are severely constrained.

\begin{figure}
\centering
\includegraphics[width = 85mm]{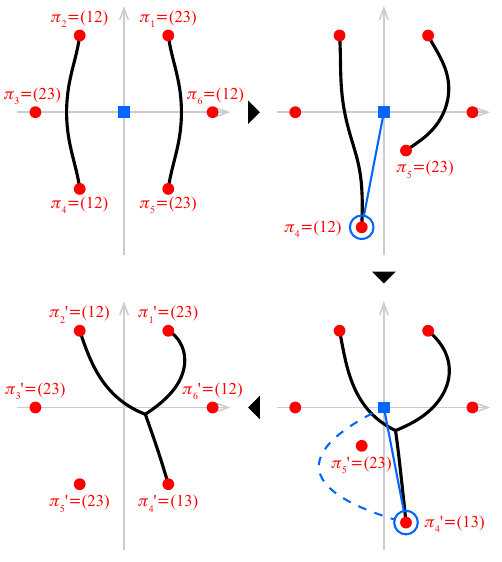}
\caption{Deformation of the model by moving the branch points on the complex $\omega$ plane leads to morphology changes in the OBC spectrum when a branch point moves through an existing branch cut.
Monodromy representations using the base point $\omega_0 = 0$ (blue squares) are given.
In the second panel, the based loop for branch point $4$ is shown.
In the third panel, $\pi_4$ is conjugated by $\pi_5$ when the previous based loop for branch point $4$ (blue dashed curve) moves past branch point $5$ (blue solid line).
Other permutation labels remain unchanged during the deformation process.
}
\label{fig:3}
\end{figure}

The information from monodromy representation can also be used to understand the various morphologies of OBC spectra \cite{tai2023zoology}. Figure \ref{fig:3} provides such an example where a Y-shaped OBC spectrum is created starting from Eq. (\ref{eq:hexagon}) by exchanging branch points $4$ and $5$ in the counterclockwise direction. When branch point $5$ moves past the based loop of branch point $4$, the permutation along that based loop is conjugated by $\pi_5$ \cite{pap2018non}, resulting in $\pi_4' = \pi_5^{-1}\pi_4\pi_5$. The based loop for branch point $5$ is unaffected with $\pi_5' = \pi_5$. The change in permutation labels is accompanied by morphological changes in the OBC spectrum where branch point $5$ moves through the left branch cut. As the branch point moves to different Riemann sheets after passing through this branch cut, the right section of the OBC spectrum terminates on the left branch cut instead of on the branch point, resulting in a differently shaped spectrum for the same location of branch points.

\begin{figure}
\centering
\includegraphics[width = 85mm]{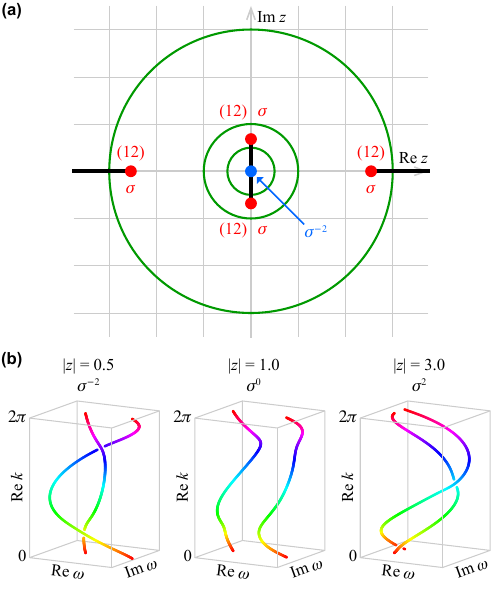}
\caption{Branch points and braid crossings in two-band models. Here $f$ is given by Eq. \ref{eq:hexagon}.
(a) Branch cuts in the $z$ plane (black solid curves). The length of each grid represents one unit of $z$. Each branch point (red points) can be associated with the braid crossing $\sigma$, and the origin (blue point) is associated with $\sigma^{-2}$. Three lops with $|z| = 0.5$ and $3$ (green dashed circles) are also shown.
(b) Braid crossings for $|z| = 0.5$, $1$ and $3$ as shown in (a). Each of these loops contains different special points that match the braid word of the spectrum. Part of the strands have been omitted from the plot to show the braiding relations.
}
\label{fig:4}
\end{figure}

{\it Braid crossings.}---Bulk band dispersions consist of complex frequencies on the Brillouin zone defined by $|z| = 1$, and are represented by braids in 3D space $(\mathrm{Re}~ \omega,~ \mathrm{Im}~ \omega,~ \mathrm{Re}~ k)$ when $k$ continuously varies from $0$ to $2\pi$ \cite{shen2018topological, wojcik2020homotopy, hu2021knots, guo2023exceptional}. Braid crossings are formed when the projections to the $(\mathrm{Re}~ \omega, \mathrm{Re}~ k)$ plane of neighboring frequency strands of the braid intersect, e.g. $\mathrm{Re}~ \omega_{(l)} = \mathrm{Re}~ \omega_{(l+1)}$, where $\omega_{(l)}$ is the roots of $f=0$ for a specific $z$ sorted in increasing real part. The braid crossings are thus located at branch cuts between the sheets defined by $\mathrm{Re}~ \omega_{(l)}$. Below, we present an analysis relating the braid behavior to the monodromy representation. This analysis is, in fact, not restricted to the Brillouin zone, and can be applied to any closed loop in the $z$ plane. As an illustration, we first focus on two-band systems, where the braid group is generated by a single braid crossing $\sigma$, defined as two strands circling around each other on the complex $\omega$ plane in the counterclockwise direction. Figure \ref{fig:4}(a) shows the branch cuts on the $z$ plane for Eq. (\ref{eq:hexagon}). Whenever the loop in the $z$ plane intersects a branch cut, $\sigma$ or its inverse is added to the corresponding braid word. To connect the braid word to the monodromy representation, we choose a based point away from all branch points as well as from $z = 0$, which may or may not be a branch point but is a pole of $\omega$ in the band structure, and decompose the loop as a sum of based loops for all branch points inside the loop and $z=0$. The overall braid word can be seen as the concatenation of braid words associated with each point. At each generic branch point, we have $\omega \sim \sqrt{z}$, and winding around a branch point counterclockwise in the $z$ plane causes the two $\omega$'s to move around each other halfway counterclockwise, resulting in a single braid crossing $\sigma$. The pole $z = 0$ corresponds to negative powers of $\sigma$. For the model of Eq. (\ref{eq:hexagon}), the two $\omega$'s scale as $z^{0}$ and $z^{-1}$ respectively near the $z=0$ pole, and is equivalent to a braid word of $\sigma^{-2}$. Although the based loops may cross other branch cuts, the conjugation by $\sigma$ does not affect the braid word as the braid group for two strands is abelian. For the same reason, the braid word does not depend on the base point or the path for the based loop, and these do not need to be specified in two-band models for the purpose of calculating braid words. As such, the number of $\sigma$ present in the braid word for any closed loop in the $z$ plane can be found by counting the number of branch points and poles inside this loop. Figure \ref{fig:4}(b) shows the braiding for three loops, $|z| = 0.5$, $1$ and $3$. The braid word in each case agrees with the number of special points within the loop.

The results can be generalized to models with three or more bands where the braid groups are not abelian. With three or more bands, the braid word can not be uniquely determined without explicitly drawing the branch cuts and specifying the based loops, but the crossing number induced by each simple branch point is $1$, independent of the based loops and permutation labels. The crossing number of the braid on a loop is the sum of crossing numbers contributed by all special points enclosed by the loop, which can be formulated as the winding number of the discriminant of $f$ on the loop (see Supplementary Information \cite{SI}). On the other hand, the overall permutation for the braid word of the loop can be obtained as a permutation product from the monodromy representation. A nontrivial permutation would imply the presence of a nontrivial braiding (see Supplementary Information \cite{SI}).

{\it Conclusion.}---We introduce the viewpoint that the band structure of a non-Hermitian system is viewed as a Riemann surface. We show that experimentally observable properties of non-Hermitian systems, including OBC spectra and braidings of the bulk bands, can be connected to the analytic and topological properties of the corresponding Riemann surface via branch cuts. Our results rely heavily on the algebraic properties derived from the band structure, including the monodromy representation and the discriminant. The generic branch points on the $z$ complex plane are also known as exceptional points where eigenfrequencies become degenerate and eigenvectors coalesce, and their significance in the classification of braiding has been well-understood \cite{hu2021knots}. Here, the relation of these special points to the discriminant allows the crossing number to be calculated without solving for individual bands. Overall, we believe the Riemann surface approach for non-Hermitian band structures will lead to new understandings of their unusual properties.

\begin{acknowledgments}
We thank Janet Zhong and Cheng Guo for their helpful discussions. This work is supported by the Simons Investigator in Physics grant from the Simons Foundation (Grant No. 827065).
\end{acknowledgments}

\appendix
\setcounter{figure}{0}
\renewcommand*{\thefigure}{S\arabic{figure}}
\setcounter{equation}{0}
\renewcommand{\theequation}{S\arabic{equation}}

\newpage
\begin{widetext}

\begin{center}
\large Supplementary Information
\end{center}

\section{Riemann-Hurwitz formula}

The Riemann-Hurwitz formula describes the relationship of Euler characteristics of two Riemann surfaces $X$ and $Y$ when $X$ is a ramified covering of $Y$ \cite{cavalieri2016riemann}. Assume that $F$ is an analytic function that maps each point on $X$ to some point on $Y$. In considering the band structure discussed in the main text, $X$ is the Riemann surface given by the characteristic polynomial $f$, each point labeled by a pair of coordinates $(z, \omega)$, $Y$ can be the $z$ plane or the $\omega$ plane, and $F$ selects one of the coordinates in $X$ as the point on $Y$. Almost all points on $Y$ have $d$ preimages, where $d$ is termed the {\it degree} of the mapping, but some special points on $Y$ will have fewer preimages. These points on $Y$ are {\it branch points} and the corresponding preimages on $X$ are {\it ramification points}. The {\it ramification index} of a point $P$ on $X$ is the number $n$ such that the map centered at $z_P$ locally looks like $F(z) \sim z^n$. Ramification points have ramification indices larger than $2$, while the other points, termed {\it unramified points}, have ramification indices $1$.

An example is the square root function given by $f: \omega - z^2 = 0$. Here $X$ is the Riemann sphere of $f$ [as it can be labeled as $(z = z,\ \omega = z^2)$ using $z$ alone, and $z$ is a Riemann sphere] and $Y$ is the Riemann sphere labeled by $\omega$ (the $\omega$ plane plus the infinity point). Here $z=0$ and $z=\infty$ are the ramification points, as each $\omega = 0$ and $\omega = \infty$ only has one distinct solution for $z$; they both have a ramification index of 2. The corresponding $\omega = 0$ and $\omega = \infty$ are branch points. We note that the ramification index is defined on $X$ rather than $Y$, as different ramification points may be mapped to the same branch point on $Y$. However, if these points are separated by adding perturbations to $F$ such that each ramified point maps to a distinct branch point, then the behavior of each branch point looks like the $n$-th root function, with $n$ the ramification index. If the structure of the monodromy representation is known, then each cycle of length $n$ in the permutation labels indicates a $n$-th root behavior, corresponding to a ramification point with ramification index $n$.

The original Riemann-Hurwitz formula \cite{hartshorne2013algebraic}, using the above concepts, can be expressed as
\begin{equation}
\chi_X = d \cdot \chi_Y - \sum_{P \in X}(k_P - 1)
\end{equation}
where $\chi_X = 2 - 2g_X$ and $\chi_Y = 2 - 2g_Y$ are the Euler characteristic of $X$ and $Y$, respectively, $g_X$ and $g_Y$ are the genus of $X$ and $Y$, respectively, and $k_P$ is the ramification index at $P$. The sum runs over all points of $X$, but only a finite number of points can be ramified, and the sum remains finite.

In the case of a mapping from the Riemann surface of $f$ to a Riemann sphere, we have $g_Y = 0$ and the above equation simplifies to
\begin{equation}
\sum_{P \in X}(k_P - 1) = 2g_X + 2d - 2.
\end{equation}
To interpret the left-hand side as the number of branch points (denoted $N_\mathrm{bp}$), we count each ramification point $P$ with ramification index $k_P$ as a total of $k_P - 1$ points with ramification index $2$ (known as {\it simple ramification}). This is justified as applying small perturbations to a higher-order ramification point leads to the lifting of degeneracy and continuously deforms the map $F$ into a new one featuring $k_P - 1$ ramification points, each with simple ramification. For band structures, we note that, while the branch points at finite $\omega$ and $z$ can always be perturbed in the way described above, the $\omega = \infty$, $z=0$, and $z=\infty$ points are related to the structure of the model, and perturbing the coupling coefficients does not always separate these points into points with simple ramifications (see below).

We discuss some types of band structures to demonstrate the application of the Riemann-Hurwitz formula.

{\bf One-band models} with $p$ right couplings and $q$ left couplings can be written as
\begin{equation}
\omega = \sum_{-p \leq s \leq q} t_{s}z^s
\end{equation}
where $t_s$ is the coupling coefficient. This can be converted into its polynomial form:
\begin{equation}
f: - z^p \omega + \sum_{0 \leq s \leq p+q} t_{s-p}z^s = 0.
\end{equation}
This always represents a Riemann sphere ($g=0$) as it can be labeled using $z$ alone. As such, the mapping from $f$ to $z$ becomes trivial with $d = 1$. This is consistent with the Riemann-Hurwitz formula that indicates no branch points can be found, and also with the fact that a meaningful braid would require at least two bands.

The mapping from $f$ to $\omega$ has a degree of $p+q$ by the fundamental theorem of algebra. The Riemann-Hurwitz formula then indicates that $N_\mathrm{bp} = 2(p+q)-2$. The branch points with a finite $\omega$ can be found by eliminating $z$ from $f=0$ and $\partial_z f = 0$, resulting in a $(p+q)$-th order discriminant in $\omega$. This accounts for $p+q$ branch points, and the other $p+q-2$ points are all at $\omega = \infty$. To see this explicitly, consider the behavior of $f$ when $\omega \rightarrow \infty$. There are $p$ solutions near $z \rightarrow 0$ where $\omega \sim z^{-p}$, and $q$ solutions near $z \rightarrow \infty$ where $\omega \sim z^q$. The two sets of solutions do not mix when $\omega$ circles around the infinity point, therefore $z=0$ and $z=\infty$ are two ramification points with ramification indices $p$ and $q$. Together they contribute the $p+q-2$ branch points at the infinity point, and the $\omega = \infty$ branch point always carries a permutation label in the conjugacy class of $(1,2,\cdots p)(p+1,p+2,\cdots p+q)$ for the monodromy representation.

{\bf The Su-Schrieffer-Heeger (SSH) model} reads
\begin{equation}
f: z \omega^2 - t_1t_2(z^2+1) - (t_1^2+t_2^2)z = 0
\end{equation}
where $t_1$ and $t_2$ are the two coupling coefficients. The equation contains a cubic term ($z\omega^2$), and $f$ can be transformed into an elliptic curve with $g=1$. The mapping from $f$ to $\omega$ has $d=2$, and the Riemann-Hurwitz formula indicates $N_\mathrm{bp} = 4$. For the Hermitian model considered here, the four branch points $t_1 + t_2$, $t_1 - t_2$, $-t_1 + t_2$ and $-t_1 - t_2$ are exactly the spectrum endpoints, with each band connecting two branch points. Here $\omega = \infty$ is not a branch point. The mapping from $f$ to $z$ also has $d=2$ and $N_\mathrm{bp} = 4$. The branch points are $z=0$, $z=-t_1/t_2$, $z=-t_2/t_1$, and $z=\infty$. All of them correspond to simple ramifications. Replacing $t_1$ and $t_2$ with non-Hermitian coupling amplitudes does not change the structure of $f$. As such, the topology of the Riemann surface of $f$, as well as the number of branch points, remain the same.

{\bf Two-band models with only nearest-neighbor couplings} have the general form of
\begin{equation}
\mathrm{det}\left(
\begin{bmatrix}
t_{11,-1}z^{-1} + t_{11,0} + t_{11,1}z & t_{12,-1}z^{-1} + t_{12,0} + t_{12,1}z\\
t_{21,-1}z^{-1} + t_{21,0} + t_{21,1}z & t_{22,-1}z^{-1} + t_{22,0} + t_{22,1}z
\end{bmatrix}
-
\begin{bmatrix}
\omega & 0\\
0 & \omega
\end{bmatrix}
\right)= 0.
\end{equation}
After some rearrangement, the above equation can be written as
\begin{equation}
f: z^2 \omega^2 + \omega \sum_{j=1}^3 A_jz^j + \sum_{j=0}^4 B_jz^j = 0.
\end{equation}
This equation can also be transformed into an elliptic curve with $g=1$. The mapping from $f$ to $\omega$ has $d=4$ and $N_\mathrm{bp} = 8$. The mapping from $f$ to $z$ has $d=2$ and $N_\mathrm{bp} = 4$. In both cases, the $\omega = \infty$ ($z = \infty$) point may or may not be a branch point, depending on the coefficients.

\section{Consistency of the OBC spectrum and the monodromy representation}

\begin{figure}
\centering
\includegraphics[width = 130mm]{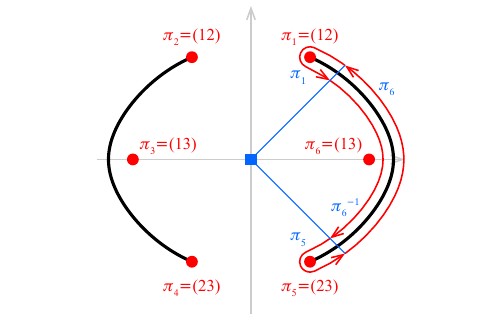}
\caption{The OBC spectrum for another model that has identical locations of branch points with Fig. 2 in the main text. The coupling coefficients are given in the next section. The consistency with the monodromy representation can be verified by decomposing the loop enclosing the branch cut into four based loops.
}
\label{fig:S0}
\end{figure}

Here we provide another model with identical branch point locations and the set of branch points in the OBC spectrum with Fig. 2 in the main text (see Fig. \ref{fig:S0}). The right half of the OBC spectrum connects branch points $1$ and $5$ but $\pi_1 \neq \pi_5$. To illustrate the connection of this OBC spectrum to the monodromy representation, the loop around this branch cut is decomposed into four segments and supplemented with straight lines from and to the base point. The bottom, right, and top segments correspond to $\pi_5$, $\pi_6$, and $\pi_1$ respectively, while the left segment corresponds to $\pi_6^{-1}$ as the arc circles around the branch point in the clockwise direction. The permutation of the entire loop can be put together as $\pi_5\pi_6\pi_1\pi_6^{-1} = 1$, confirming the consistency. The labels being different at the two ends of a branch cut is a consequence of the dependency of the labels on the base point and the based loops chosen. It also demonstrates the fact that the labels in the monodromy representation are not Riemann sheet indices; the $z_\alpha$ designation is always continuous with $\omega$, and their association with $z_{(l)}$ changes precisely when the based loop crosses a branch cut. 

\section{Construction of non-Hermitian models with preset branch points}

Models with preset branch point locations may be constructed through algebraic or numerical approaches. The number of branch points should be consistent with the Riemann-Hurwitz formula above. The number of degrees of freedom from the undetermined coefficients of $f$ may be more or less than the conditions required from the branch points. If the model has more degrees of freedom than the number of branch points, there can be infinitely many OBC spectra for the same set of branch points. On the other hand, if the model has fewer degrees of freedom than the number of branch points, additional algebraic constraints on the branch point locations will appear.

As an example, we study the following polynomial used in the main text:
\begin{equation}
f: z \omega^2 + \omega \sum_{s=0}^{2}A_s z^s + \sum_{s=0}^{3}B_s z^s = 0.
\end{equation}
This polynomial is third-order in $z$, and the discriminant for a general third-order equation $C_3z^3 + C_2z^2 + C_1z + C_0 = 0$ can be found as \cite{gelfand1994discriminants}
\begin{equation}
\Delta_z(f) = C_1^2 C_2^2 - 4 C_0 C_2^3 - 4 C_1^3 C_3 + 18 C_0C_1C_2C_3 - 27 C_0^2C_3^2.
\end{equation}

Expanding $f$ and comparing with the form of the third-order equation, we find that
\begin{align}
C_3 &= B_3,\\
C_2 &= B_2 + A_2 \omega,\\
C_1 &= B_1 + A_1 \omega + \omega^2, \\
C_0 &= B_0 + A_0 \omega.
\end{align}
By substituting these coefficients into $\Delta_z(f) = 0$, we obtain a sixth-order equation for $\omega$ in the form of $\sum_{s=0}^6 D_s \omega^s = 0$ where each $D_s$ is a polynomial of $A_s$ and $B_s$. The six solutions of this equation give the locations of the branch points on the $\omega$ plane. For the branch points to appear at the six roots of unity, the equation should match the form $\omega^6 - 1 = 0$. This requires that
\begin{equation}
D_1 = D_2 = D_3 = D_4 = D_5 = 0,
\end{equation}
\begin{equation}
D_6 = -D_0 \neq 0.
\end{equation}
These equations impose 6 constraints on the 7 coefficients $A_{0,1,2}$ and $B_{0,1,2,3}$. However, performing gauge transformations on $z$ (i.e., $A_s \rightarrow \lambda^{s-1} A_s$ and $B_s \rightarrow \lambda^{s-1} B_s$) does not change the locations of the branch points or the OBC spectrum, therefore removing 1 degree of freedom from the coefficients for the purpose of calculating OBC spectra. As such, an arrangement of the six branch point locations corresponds to finitely many OBC spectra, each one represented by a set of equivalent $A_{0,1,2}$ and $B_{0,1,2,3}$ coefficients.

There are a total of 120 solutions (counted with multiplicities) for the coefficients for a general configuration of branch points; all of them are distinct if the branch points are located at the six roots of unity. Equation (4) in the main text is one such solution, with
\begin{equation}
A_2 = \frac{1}{\sqrt[3]{2}},\ A_1 = 0,\ A_0 = -\frac{1}{\sqrt[3]{2}},
\end{equation}
\begin{equation}
B_3 = \frac{1}{3\sqrt[3]{4}},\ B_2 = 0,\ B_1 = -\frac{1}{\sqrt[3]{4}},\ B_0 = 0.
\end{equation}
For comparison, the model used in Fig. \ref{fig:S0} in the main text has the following parameters:
\begin{equation}
A_2 = \frac{1}{\sqrt[3]{2}},\ A_1 = 0,\ A_0 = -\frac{\sqrt[3]{4}}{\sqrt{3}}\cos\frac{\pi}{18},
\end{equation}
\begin{equation}
B_3 = -\frac{\sqrt[3]{2}}{3\sqrt{3}}\cos\frac{\pi}{18},\ B_2 = 0,\ B_1 = -\sqrt[3]{2}\cos\frac{\pi}{9},\ B_0 = 0.
\end{equation}
The model used in the last panel of Fig. 3 in the main text has the following parameters:
\begin{equation}
A_2 = 0.79370,\ A_1 = -0.45129 - 0.49865i,\ A_0 = -0.45426 -0.36110i,
\end{equation}
\begin{equation}
B_3 = 0.16097 +0.00821i,\ B_2 = -0.16002 -0.23901i,
\end{equation}
\begin{equation}
B_1 = -0.26508 -0.06631i,\ B_0 = -0.06357 +0.39112i.
\end{equation}
We note again that these parameters should be interpreted as a representative element in the equivalence class of gauge transformations $A_s \rightarrow \lambda^{s-1} A_s$ and $B_s \rightarrow \lambda^{s-1} B_s$.

To realize the above $f$ as two-band models, We expand the Hamiltonian in the main text as a characteristic equation:
\begin{equation}
\begin{vmatrix}
a_1 z + a_0 + a_{-1}z^{-1} - \omega & u + v z^{-1} \\ 
v z + u & b_1 z + b_0 - \omega
\end{vmatrix}
=0,
\end{equation}
which can be written equivalently as
\begin{align}
z \omega^2 + \omega\left[ (-a_1-b_1)z^2 + (-a_0-b_0)z - a_{-1} \right] & \nonumber\\
+ \left[a_1 b_1 z^3 + (a_1b_0 + a_0b_1 - uv)z^2 + (a_0b_0+a_{-1}b_1-u^2-v^2)z + (a_{-1}b_0-uv) \right] &= 0.
\end{align}
This can be compared with the coefficients above and the coupling amplitudes can be solved. For example, for Eq. (4) in the main text, the solutions for coupling amplitudes are
\begin{equation}
a_{-1} = \frac{1}{\sqrt[3]{2}},\ a_0 = b_0 = 0,
\end{equation}
\begin{equation}
a_1 = -\frac{1}{\sqrt[3]{2}} \frac{\exp(\pm\pi i/6)}{\sqrt{3}},\ b_1 = a_1^*,
\end{equation}
\begin{equation}
\{u,v\} = \left\{0, \pm\sqrt{\frac{-a_1}{\sqrt[3]{2}}}\right\}.
\end{equation}
Each $\pm$ can independently take $+$ or $-$ values and $u$ and $v$ can be interchanged, giving a total of 8 solutions. It can be directly verified that all the above models have the same continuous part of the OBC spectra shown in Fig. 2 in the main text. The existence of edge states, however, is different between the models, as the bulk band does not uniquely determine how the chain is terminated at the boundary.

Models carrying different monodromy representations can also be converted to each other through deformation. In the case where branch point locations lead to finitely many OBC spectra, moving the branch points on the complex $\omega$ plane can be accomplished by continuously changing the coefficients in $f$. To track the parameters in $f$ while the branch point is moving, a standard optimization algorithm can be used to minimize the distance between the current branch points of $f$ and the intended locations of branch points along the path. This can also be used to generate models with arbitrary branch point locations and monodromy representations by starting from an algebraically more traceable one. Models in Fig. 3 in the main text are found with this method. The model used in the second panel of Fig. 3 in the main text has the following parameters:
\begin{equation}
A_2 = 0.79370,\ A_1 = -0.18811 -0.03995i,\ A_0 = -0.84452 -0.08281i,
\end{equation}
\begin{equation}
B_3 = 0.18470 +0.02750i,\ B_2 = -0.10504 -0.06318i,
\end{equation}
\begin{equation}
B_1 = -0.50980 -0.17716i,\ B_0 = 0.2387 +0.19376i.
\end{equation}
The model used in the third panel of Fig. 3 in the main text has the following parameters:
\begin{equation}
A_2 = 0.79370,\ A_1 = -0.37619 -0.18098i,\ A_0 = -0.76125 -0.31237i,
\end{equation}
\begin{equation}
B_3 = 0.16482 +0.01815i,\ B_2 = -0.14531 -0.13348i,
\end{equation}
\begin{equation}
B_1 = -0.34604 -0.17385i,\ B_0 = 0.12461 +0.42164i.
\end{equation}

\section{Crossing number as the winding number of the discriminant}

The discriminant of $f$ as a polynomial in $\omega$ is defined as
\begin{equation}
    \Delta_\omega (f) = (-1)^{r(r-1)/2} D_r^{2r-2} \prod_{m\neq n} (\omega_m - \omega_n) = D_r^{2r-2} \prod_{m>n} (\omega_m - \omega_n)^2
\end{equation}
where $r$ is the degree of $f$ in $\omega$ (also the number of bands), $D_r$ is the coefficient for $\omega^r$ in $f$, and $\omega_m$ are the $r$ roots of $f=0$. While the roots may not be written as algebraic expressions, the discriminant itself can be constructed as polynomials of the coefficients of $f$ and therefore efficiently computable. As a zero discriminant implies some form of degeneracy in the system (e.g. an exceptional point), it has been applied to classifying higher-dimensional band structures \cite{wojcik2020homotopy, yang2021fermion}. The form of $\Delta_\omega (f)$ can track the relative locations of $\omega_s$ as they wind around each other on the complex plane; the winding number of each individual factor has been studied as the ``inter-band energy vorticity'' \cite{leykam2017edge, shen2018topological}. Below, we directly prove that the crossing number of a braid is given by the winding number of the discriminant as a function of $z$ around $\Delta_\omega (f)=0$. This will be done by sequentially establishing that a braid diagram whose projection onto $\mathrm{Re}~ \omega$ has no intersections has $0$ crossing number and $0$ winding number; a braid diagram with a single intersection in the projection has $\pm 1$ crossing number and correspondingly $\pm 1$ winding number; and a general braid diagram can be reduced to the above two cases.

Assume that there are no $k$ values such that $\mathrm{Re}~ \omega_m(k) = \mathrm{Re}~ \omega_n(k)$. Since $\omega_m$ is continuous along $k \in [0, 2\pi]$, the $\omega_m$ can be ordered by their real part by assigning $\omega_m = \omega_{(m)}$, where $\omega_{(l)}$ is the roots of $f=0$ for a specific $z$ sorted in increasing real part, and this assignment will not change as $k$ varies. Each factor in the discriminant $\omega_m - \omega_n$ with $m > n$ always has a positive real part, and the argument of $(\omega_m - \omega_n)^2$ lies in the range of $(-\pi, \pi)$ as $k$ varies. Since $(\omega_m - \omega_n)^2$ never passes through the negative real axis, it can not wind around the zero point, and its winding number is $0$. The winding number of $\prod_{m>n} (\omega_m - \omega_n)^2$ is the sum of the winding number of each factor, and is also $0$.

Now assume that there is a point $k=k^*$ where two of the $\mathrm{Re}~ \omega_m(k)$ are equal and no crossing occurs elsewhere. We again assign $\omega_m = \omega_{(m)}$ at $k = 0$ and continuously trace the $\omega_m$ along the strands. Denote the strands with equal real parts at $k^*$ as $\omega_\mu$ and $\omega_{\mu+1}$. When $k > k^*$, these two strands will exchange their ordering of real parts, and we have $\omega_\mu = \omega_{(\mu+1)}$ and $\omega_{\mu+1} = \omega_{(\mu)}$. The factor $(\omega_{\mu+1} - \omega_{\mu})^2$ will cross the negative real axis at $k^*$ and move around the origin once. If $\mathrm{Im}~ \omega_\mu < \mathrm{Im}~ \omega_{\mu+1}$ at $k = k^*$, the argument of $(\omega_{\mu+1} - \omega_{\mu})^2$ increases near $k^*$, and the winding number is $1$, while $\omega_\mu$ and $\omega_{\mu+1}$ circles around each other counterclockwise and have a crossing number of $1$. Similarly, if $\mathrm{Im}~ \omega_\mu > \mathrm{Im}~ \omega_{\mu+1}$ at $k = k^*$, both the winding number of $(\omega_{\mu+1} - \omega_{\mu})^2$ and the crossing number is $-1$. All other factors from $\prod_{m>n} (\omega_m - \omega_n)^2$ do not cross the negative real axis and do not contribute to the winding number.

Finally, for an arbitrary braid diagram, we can segment the range of $k$ such that each section contains at most one crossing. If there are two or more crossings at the same $k$, the pair of strands involved must be different for each crossing. The crossings can therefore be slightly perturbed and moved to a different $k$, which does not affect either the crossing number or the winding number. Each segment can be converted to a closed braid by connecting each $\omega_{(m)}$ to the location of $\omega_{(m)}(k=0)$ with a straight line on the $\omega$ plane. Since the connection preserves the ordering of real parts, the projections of these straight lines do not cross, and the winding number for each segment equals the crossing number. The original braid diagram can be recovered by joining the different segments together. The added straight lines cancel out each other, and the total winding number of $\prod_{m>n} (\omega_m - \omega_n)^2$ gives the crossing number of the entire braid diagram.

The above proof only requires $f$ to be a polynomial in $\omega$ rather than $z$, and can be extended to lattice models with long-range couplings as long as $D_r \neq 0$ everywhere along the strands. We note that we have not considered the winding number of the term $D_r^{2r-2}$. In order for the winding number of $\Delta_\omega (f)$ to agree with $\prod_{m>n} (\omega_m - \omega_n)^2$, the winding number of $D_r$ must be $0$. This can be readily achieved if $D_r$ is a constant. When calculating the characteristic polynomial of $H$ from $\mathrm{det}(H-\omega\mathbb{I})=0$, the coefficient of the leading term in $\omega$ will be $\pm 1$ and automatically satisfy the condition. If instead $f$ is known, then dividing $D_r$ throughout $f$ results in an equivalent polynomial in $\omega$ that can be used for the winding number calculations. 

\section{Braidings and permutations for multiband systems}

To demonstrate the relations between monodromy representations, permutation labels, and braidings in multiband systems, we consider the following three-band example:
\begin{equation}
\omega^3 - (2z^2-z^{-2})\omega + (z^2+z^{-2}) = 0.
\end{equation}
We define $\sigma_\mu$ as the crossing where $\mathrm{Re}~ \omega_{(\mu)} = \mathrm{Re}~ \omega_{(\mu+1)}$ and the two strands circle around each other in the counterclockwise direction. With this notation, the braiding of frequencies on the Brillouin zone (BZ) is given by $\sigma_1\sigma_2^{-1}\sigma_1\sigma_2^{-1}$ [Fig. \ref{fig:S1}(a)] and forms a figure-eight knot when all strands are joined.

The branch cuts on the $z$ plane associated with $\sigma_1$ and $\sigma_2$ are shown in Fig. \ref{fig:S1}(b). These branch cuts begin from the branch points and end at either $z=0$ or $z=\infty$. Crossing a branch cut on the $z$ plane adds a corresponding braid crossing to the braid word. A positive crossing $\sigma_\mu$ is added if the crossing direction circles counterclockwise around the branch point it is connected to, and a negative crossing $\sigma_\mu^{-1}$ is added if clockwise. To relate the braiding to the monodromy representation, we list the permutation label and the braid word along each based loop below [the numbering of branch points are given in Fig. \ref{fig:S1}(b)]:
\begin{align*}
\pi_{ 1} = (12) & \leftarrow \sigma_1
&
\pi_{ 2} = (23) & \leftarrow \sigma_2
\\
\pi_{ 3} = (12) & \leftarrow \sigma_1
&
\pi_{ 4} = (23) & \leftarrow \sigma_2
\\
\pi_{ 5} = (12) & \leftarrow \sigma_1
&
\pi_{ 6} = (13) & \leftarrow \sigma_1^{-1}\sigma_2\sigma_1
\\
\pi_{ 7} = (23) & \leftarrow \sigma_1^{-1}\sigma_2^{-1}\sigma_1\sigma_2\sigma_1
&
\pi_{ 8} = (12) & \leftarrow \sigma_1\sigma_2\sigma_1\sigma_2\sigma_1^{-1}\sigma_2^{-1}\sigma_1^{-1}
\\
\pi_{ 9} = (23) & \leftarrow \sigma_1\sigma_2\sigma_1\sigma_2^{-1}\sigma_1^{-1}
&
\pi_{10} = (12) & \leftarrow \sigma_1\sigma_2\sigma_1\sigma_2\sigma_1^{-1}\sigma_2^{-1}\sigma_1^{-1}
\\
\pi_{11} = (23) & \leftarrow \sigma_1\sigma_2\sigma_1\sigma_2^{-1}\sigma_1^{-1}
&
\pi_{12} = (13) & \leftarrow \sigma_1\sigma_2\sigma_1^{-1}
\end{align*}
We observe that the portion of the based loop close to the branch point will intersect the branch cut originating from this branch point, leading to a single positive crossing $\sigma_\mu$. However, as the straight line segment in the based loop may cross other branch cuts, the overall braid word can be any element in the conjugation class of $\sigma_\mu$. While the exact braid word can not be recovered from the monodromy representation alone, as it does not concern the relative locations when moving from the base point to the vicinity of the branch point, the permutation label can be implied from the braid word by mapping the braid group to the permutation group, i.e. $\sigma_1 \rightarrow (12)$ and $\sigma_2 \rightarrow (23)$. The crossing order induced by each simple branch point is $1$, independent of the specific permutation label. The $z=0$ point is not a branch point in this case, but a based loop around it carries a nontrivial braid word $\sigma_1^{-1}\sigma_2^{-1}\sigma_1^{-2}\sigma_2^{-1}\sigma_1^{-1}$. By decomposing the BZ as the sum of based loops circling all branch points inside the BZ as well as $z=0$, the permutation along the BZ is obtained as the ordered product of all permutation labels from the based loops and reads
\begin{equation}
\pi_{1}\pi_{3}\pi_{5}\pi_{7}\pi_{9}\pi_{11} = (132).
\end{equation}
Similarly, the entire braid word is obtained as the ordered product of all braid words from the based loops and is $\sigma_1^2\sigma_2^{-1}\sigma_1\sigma_2^{-1}\sigma_1^{-1}$, agreeing with Fig. \ref{fig:S1}(a) up to conjugation. The corresponding permutation for the braid word is consistent with the permutation obtained from monodromy representations, and a nontrivial permutation would imply the presence of a nontrivial braid.

Finally, we plot the discriminant $\Delta_\omega(f)$ along the BZ in Fig. \ref{fig:S1}(c). Here the discriminant does not wind around $\Delta_\omega = 0$, and the winding number $0$ agrees with the crossing number of the braid word for the BZ.

\begin{figure}
\centering
\includegraphics[width = 170mm]{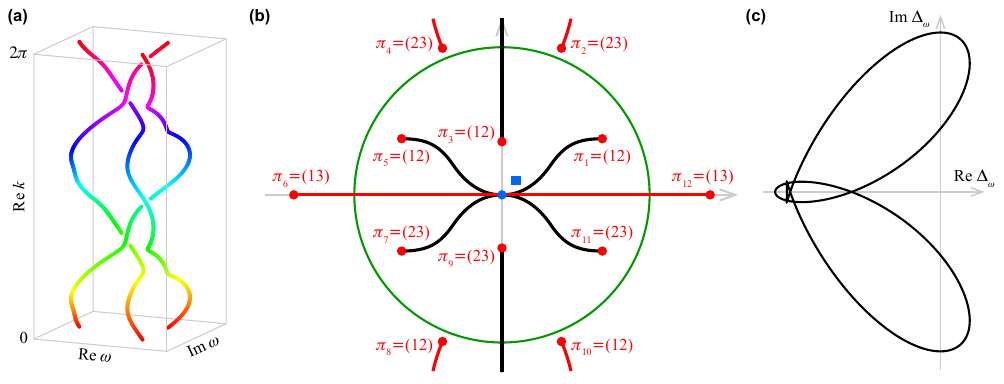}
\caption{Branch points and braid crossings of a three-band model.
(a) Braid crossings for the three-band model $\omega^3 - (2z^2-z^{-2})\omega + (z^2+z^{-2}) = 0$. 
(b) Branch cuts in the $z$ plane for the three-band model. The branch cuts corresponding to $\sigma_1$ and $\sigma_2$ have been colored black and red respectively for distinction. Branch points (red points) are numbered and their permutation label is given (the base point is marked as the blue square). The origin is marked by the blue point. The Brillouin zone is marked by the green circle.
(c) Trajectory of the discriminant $\Delta_\omega$ in the complex plane, with a winding number of $0$.
}
\label{fig:S1}
\end{figure}

\end{widetext}

%

\end{document}